\documentclass[lettersize,journal]{IEEEtran}
\usepackage{amsmath,amsfonts}
\usepackage{algorithmic}
\usepackage{array}
\usepackage[caption=false,font=normalsize,labelfont=sf,textfont=sf]{subfig}
\usepackage{textcomp}
\usepackage{stfloats}
\usepackage{url}
\usepackage{verbatim}
\usepackage{graphicx}

\usepackage{xcolor}

\usepackage[linesnumbered,lined,commentsnumbered]{algorithm2e}%
\usepackage{hyperref}
\hypersetup{hidelinks}

\SetAlCapSty{footnotesize}

\usepackage[backend=bibtex,style=ieee,sorting=none]{biblatex}
\bibliography{bibl}  
\usepackage{makecell}
\usepackage{setspace}
\usepackage{arydshln, booktabs}

\usepackage{authblk}

\usepackage{soul}
\setuldepth{0.1}
\setul{}{0.01 pt}

\usepackage[T1]{fontenc}
\usepackage[outline]{contour}
\contourlength{0.1pt}
\contournumber{10}%

\usepackage{flushend}

\usepackage[%
    binary-units=true,
    prefixes-as-symbols=false,
    ]{siunitx}

\newcommand{\orcidi}[1]{\href{https://orcid.org/#1}{\includegraphics[width=8pt]{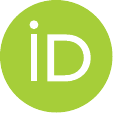}}}

\def\BibTeX{{\rm B\kern-.05em{\sc i\kern-.025em b}\kern-.08em
    T\kern-.1667em\lower.7ex\hbox{E}\kern-.125emX}}
\usepackage{balance}
\begin{document}
\title{Enthuse: Efficient Adaptable High-throughput\\ Streaming Aggregation Engines}

\author{Philippos Papaphilippou, Wayne Luk}

\author{Philippos Papaphilippou
\IEEEcompsocitemizethanks{\IEEEcompsocthanksitem Philippos Papaphilippou is with the School of Electronics and Computer Science, University of Southampton, United Kingdom
(E-mail: pp1d24@soton.ac.uk). 
}\orcidi{0000-0002-7452-7150},~\IEEEmembership{Member,~IEEE}, 
Wayne Luk
\IEEEcompsocitemizethanks{\IEEEcompsocthanksitem Wayne Luk is with the Department of Computer Science, Imperial College London, United Kingdom
(E-mail: w.luk@imperial.ac.uk). 
}\orcidi{0000-0002-6750-927X},~\IEEEmembership{Fellow,~IEEE}}

\maketitle

\begin{abstract}

Aggregation queries are a series of computationally-demanding analytics operations on counted, grouped or time series %
data. They include tasks such as summation or finding the median among the items of the same group, and within a specified number of the last observed tuples for sliding window aggregation (SWAG). They have a wide range of applications including 
database analytics, operating systems, bank security and medical sensors.
Existing challenges include %
 the hardware complexity that comes with efficiently handling per-group states using hash-based approaches.
This paper presents Enthuse, an adaptable pipeline for calculating a wide range of aggregation queries with high throughput. %
It is then adapted for SWAG and achieves up to 476x speedup over the CPU core of the same platform. %
It %
achieves unparalleled levels of performance and functionality such as a throughput of 1 GT/s  on our setup for SWAG without groups, and more advanced operators with up to 4x the window sizes than the state-of-the-art with groups as an approximation for SWAG featuring per-group windows using a fraction of the resources and no DRAM.

\end{abstract}


\section{Introduction}\label{intro}

\IEEEPARstart{A}{ggregation} is the task of combining raw data to produce more useful and summarised knowledge. It appears in many disciplines, %
especially relating to distributed and parallel programming models, such as MapReduce \cite{Tangwongsan2018}, where multiple workers/nodes produce a partial result that needs aggregating. The worker notion is also popular in FPGA designs \cite{owaida2017centaur}. Equivalently, an established software design pattern is the fork-join model, which is a prominent concurrency pattern %
with processes or threads calculating partial results \cite{de2014fork}.

An emerging need in data analytics is the use of programming models and related infrastructure to be able to benefit from the increasingly-wide memories and datapaths \cite{prra}. A more traditional alternative is to have multiple workers and ports, as with FPGA-based database designs with multi-threading-equivalent arrangements. However, since the memory access pattern is highly-affected by this design choice, the hardware complexity is also affected, such as by %
co-designing cache or hash structures on the programmable logic. 
In the software world, this has already been reflected in the prominence of the key-value store paradigm, where the simplified storage model can enable higher throughput \cite{zhang2015mega}. 

FPGAs are more than capable of accelerating complex tasks such as aggregation while utilising wide datapaths. This can be achieved ``naturally'' by introducing %
state machines that perform fine-grain parallelisation, or with multiple-workers on buffered data. In order to simplify the design of such high-throughput %
operators, the hardware programmer can implement modular designs based on optimised building blocks. %

One typical %
aggregation application is the \textit{group-by aggregate} operator for relational databases. It provides a summary of rows belonging to (having) the same \textit{group ID}. %
An example SQL (structured query language) snippet is provided in algorithm \ref{al1}, where a sum operation is applied to all keys (\texttt{table0.key2}) of the same group (\texttt{table0.key1}). Other than summation, %
SQL aggregate functions include  
\textit{minimum, maximum, average} and \textit{count}. These are sometimes combined with more advanced options such as for calculating the distinct key count per \textit{group ID}.%

\begin{algorithm}[h]
\vspace{-0.4em}
\LinesNumbered
\SetSideCommentRight

\texttt{\textbf{\textcolor{gray}{SELECT}} table0.key1, \textbf{\textcolor{gray}{sum}}(table0.key2)}

\texttt{\textbf{\textcolor{gray}{FROM}} table0}

\texttt{\textbf{\textcolor{gray}{GROUP BY}} table0.key1}

\texttt{\textbf{\textcolor{gray}{ORDER BY}} table0.key1}

\vspace{0.5em}
\caption{\footnotesize SQL example of group-by-aggregate for summation. %
}\label{al1}
\vspace{-0.4em}
\end{algorithm}

A similar category of queries is %
sliding window aggregation (SWAG) \cite{Tangwongsan2018}. 
A time domain is associated with the data, %
as with queries specifying a time range %
to form windows %
for aggregation. %
Due to the fact that the window formation is applied on-the-fly, near where the data is gathered, %
the stream is considered partly sorted with regard to accompanying timestamps or the implied order of the data. %
SWAG usually yields a random memory access pattern to calculate, including for fine-tuned CPU algorithms %
\cite{theodorakis2018hammer}. An example is when finding the median, as even with FPGAs, current designs require caching of larger states \cite{geethakumari2023stream}. This complicates the design by increasing the on-chip and off-chip memory requirements for supporting/approaching the worst case (as with multiple inputs hashing to the same slot for hashsets).

\begin{figure*}[h!]
\centering
\includegraphics[width=0.93\textwidth, trim=0 6 0 2]{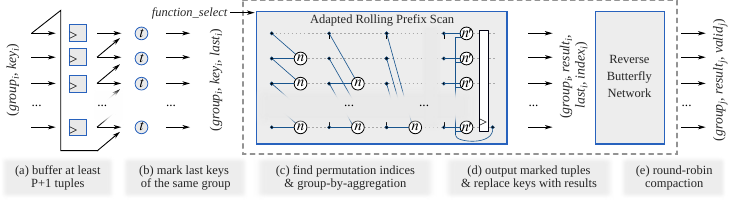} %
\caption{Five-step pipeline of the proposed aggregation pipeline of Enthuse. %
}\label{figengine}
\vspace{-0.6 em}
\end{figure*}

This paper presents Enthuse, a %
novel design for aggregation that integrates computation inside scalable structures such as the prefix scan and the butterfly network that process multiple elements per-cycle (high-throughput). %
The idea behind this design is the \emph{exploitation of those structures} originally used for highly-optimised compaction \cite{prra} to \emph{simultaneously append aggregation information}, leading to minimal resources (efficiency).
The resulting base pipeline supports a plethora of operators and use cases (adaptable), as  with the proposed
adaptations for performing SWAG queries, by %
accompanying an on-chip sorter-cardinality counter. %
The resulting SWAG engines are highly memory-efficient to the point that they achieves similar functionality to highly-complex competing designs without their limiting requirement for an off-chip DRAM %
or compression. %
Conveniently, all logical states fit entirely inside on-chip memories such as block RAM (BRAM). 
For applications such as SWAG, the concept behind its advantageous system behaviour is the replacement of the need for large indexed or hashed structures %
\textit{by offloading the design complexity to sorting}, as it eliminates fragmentation originally caused by maintaining per-group information.
FPGA-based high-throughput sorters are thoroughly studied in literature \cite{saitoh2018high, zuluaga2016streaming, qiao2022topsort, song2016parallel}, but in order to seamlessly support the median operator in our SWAG engine, innovative lightweight algorithmic modifications are introduced. %

\vspace{0.3em}

The contributions of Enthuse are as follows: 
\begin{itemize}
\item An efficient %
adaptable group-by-aggregate engine for sorted streams %
that processes multiple inputs per cycle. 
\item A state-of-the-art DRAM-less sliding window aggregation %
engine optimised for the single-window use case. %
\item A series of modifications for a high-throughput sorter %
so that it can append group cardinality to each tuple for the purposes of supporting the SWAG median operator. %
\item Open-source system-based and out-of-context evaluation of the proposed designs.

\end{itemize}

This journal paper extends the similarly-named conference poster ``Efficient Adaptable Streaming Aggregation Engine'' \cite{11008955} especially with adaptations in novel sliding-window aggregation engines (section \ref{swagsec}) and a comprehensive theoretical and system-level evaluation (section \ref{eval}).

\section{Enthuse -- aggregation framework}

In this section, we propose a novel pipeline as a building block %
for performing a range of commonly used analytics operators. %
This engine is later adapted to perform a wider selection of queries with sliding windows %
 (section \ref{swage}).

The main idea behind Enthuse is that \emph{the pre-existing prefix-scan topology} of a sparsity-removal circuit can be \emph{exploited and reused} to also calculate and append aggregation information to the data, before filtering redundant tuples. The focus of this pipeline is high-throughput, as it is always able to process up to \(P\) elements per cycle. It is inspired by the PRRA architecture \cite{prra}, where a rolling prefix scan is used to calculate the permutation indices for the permutation network (reverse butterfly) to achieve stream compaction (see figure \ref{figprra0}, left).

\subsection{Supported functionality}\label{subfu}

The accepted input stream can include a group associated with each value (key-value pairs), or without, as with the simpler task of summarising plain integers. The engine expects a sorted 
stream consisting of (\(group_i, key_i\)) tuples, arriving in batches of size \(P\) per cycle. \(P\) can be adjusted at synthesis-time to accommodate wide datapaths accordingly (elaborated in section \ref{scal}). %
The input %
is assumed dense, since the engine either receives \(P\) or 0 tuples in a cycle, %
according to other latencies at the system-level. %
The \(group_i\) part is optional, as it can be configured to accept number streams instead, which is equivalent to associating all values to the same group.

In terms of operators, due to the prefix scan component, the pipeline supports (a superset of) associative operators, within the context of the values belonging to the same group. That is \(a\oplus (b \oplus c) = (a \oplus b) \oplus c  \ \forall\  a,b,c \in S\), where \(S\) is the value set, such as all possible 32-bit integers. %
The studied operators include \textit{minimum}, \textit{maximum} and \textit{summation}, \textit{count}. It also supports combinations of the above, such as the \textit{average}, by simultaneously calculating  \textit{summation} and \textit{count}, and then combining the results near its output. When the input is entirely sorted including by-the-value, this extends its applicability, such as for performing \textit{distinct count} (essentially an associative operation when the input is sorted, see step \((c)\)). %

This design is a non-blocking building block working as a pipeline, i.e. it never asserts backpressure on the stream, as with the PRRA architecture. On the contrary, aggregation implies summarisation, hence the need to remove a newly generated intermediate sparsity with the compaction functionality of the permutation network \cite{prra}. %
For most aggregation tasks, this will result in only the last tuple per group being considered useful, since it is repurposed to carry all the relevant accumulated information, while the remaining aforementioned ``redundant'' tuples may carry partial results. %

\subsection{Pipeline steps}\label{gba}

Figure \ref{figengine} introduces the architecture of Enthuse, %
a high-throughput pipeline for calculating a wide range of aggregation tasks including group-by. As an adaptable framework, it consists of a series of generalisable steps: %

\vspace{0.2em}
\paragraph*{Step (a) Buffering} %
this step is responsible for buffering up to one extra batch of size \(P\), so that the entities (nodes) \(t\) of the next step (b) are able to distinguish and mark each last tuple per \(group\). %

\vspace{0.2em}
\paragraph*{Step (b) Marking last keys} essentially, this marking is done by setting the tuple's \(last\) bit high, representing the last tuple for each group. These last tuples are important, because they will be those that will carry the aggregate results, as they will be able to ``see'' all previous values of their \(group\) before being extracted. Thus, these bits will also be used as valid-bits for the reverse butterfly network, to denote a valid tuple, also containing the summarised information after the end of the pipeline. %

\vspace{0.2em}
\paragraph*{Step (c) Prefix-scans}the entities named \(n\) are distributed inside the adapted %
prefix scan to calculate the permutation index of the tuples to achieve the round-robin effect by the reverse butterfly network, as with the PRRA \cite{prra}. %
At the same time they also operate on the \(key\) fields, which later becomes the \(result\) field containing the aggregation summary per group. For example, if the query function is to find the sum of all the keys for each of the group IDs, the entities \(n\) not only perform a cumulative sum for the reverse butterfly \(index\), but also for the key sum per group, until the last entry of the same group is processed to carry the full sum of the group. %
An equality check on the group field with the previous entry (\(group_i\stackrel{?}{=}group_{i-1}\)) is enough for knowing where not to propagate the previous result. Using \textit{sum} as an example, the start of a new \(group\) would leave the current \(key\) intact instead of adding the result of the last tuple.

In order to also support the \textit{distinct count} operator (numerical example illustrated in figure \ref{figprra0}, right), 
the entities \(n\) also need to follow a simple distributed algorithm: %
if the propagated minimum from the input node of the larger key differs from the smaller key, then the distinct counts are coming from disjoint sets, hence \(n\) adopts their sum. If the minimum is equal to the smaller key of the two, then  \(n\) adopts this sum of distinct counts, but subtracts 1 for taking care of the double counting of 
the common key. %

\vspace{0.2em}
\paragraph*{Step (d) Rollover and summarisation} the entities \(n'\) are special versions of the entities \(n\) and their main
 purpose is to output the originally-marked tuples as valid to the permutation network, while replacing their key field with the result. The main difference with \(n\) is that sometimes the partial result information cannot be contained in the \(key\) field, and they behave as entrypoints for taking additional signals into account for these final results. 

For instance, if \(function\_select\) indicates that the operation is to find the average values per group, the entities \(n\) operate the same as in the sum operation. However, it is the entity \(n'\) that will divide the result by the corresponding group tuple count (cardinality). In such a case, the entities \(n\) also include internal signals for counting the number of tuples in the group in a similar fashion to the sum operation, but adding 1 instead of the key. A related difference is with the count signals of \(n'\), since \(n'\) are responsible for the rolling effect, the propagated count of the bottom \(n'\) is able to count beyond \(P\) elements, e.g. by having a 32-bit width. All \(n'\) elements use this value plus the \((log_2P+1)\)-bit wide count from the prefix sum to provide the final 32-bit count in the result, when it is requested such as from a count query. According to the complexity of the supported operators, additional signals may be carried in the network, such as the minimum, %
that becomes of use when calculating the \textit{distinct count} per group ID. %

\vspace{0.2em}
\paragraph*{Step (e) Round-robin compaction} finally, the reverse butterfly network %
permutes the batch of results to achieve the round-robin effect in parallel (see figure \ref{figprra0}, left), and is inline with \cite{prra}. The reverse butterfly network is used as an optimal-complexity switch, though other permutation networks can also be used. The \(last\) flag is now renamed to a \(valid\) bit, as it now indicates that a port has a valid result.

\subsection{Numerical example}

\begin{figure}[h!]
\centering
\includegraphics[width=0.49\textwidth, trim=0 4 0 4]{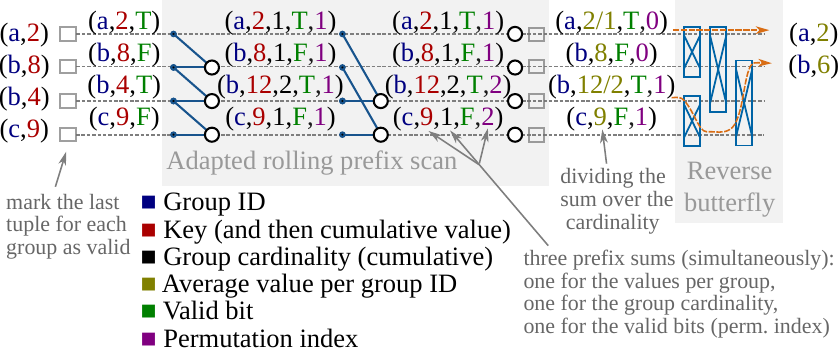} %
\caption{Colour-coded numerical example: finding the \textit{average} per group. The entries in the colour legend appear in the order of usage within each tuple.\\ \textit{See \url{https://philippos.info/groupby_visual} for an animated version online.}}\label{fignumex1}
\vspace{-0.3 em}
\end{figure}

Figure \ref{fignumex1} introduces a colour-coded numerical example %
on how a single batch of \(P\) tuples is processed through the pipeline for the \textit{sum} operator, and \(P=4\). As input, the 4 tuples have group IDs ranging from \{a,b,c\}, and they are sorted. Whether last tuples-per-group exist in the current batch is known for %
the first \(P-1=3\) tuples. %

\begin{figure*}[h!]
\centering
\includegraphics[width=0.94\textwidth, trim=0 6 0 5]{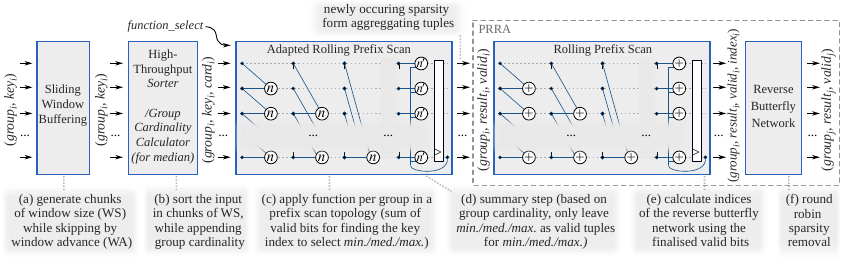} %
\caption{High-throughput adaptable sliding window aggregation %
using the six-step pipeline of EnthuseSWAG.
}\label{figswengine}
\vspace{-0.6 em}
\end{figure*}

In this instance, (a,2) and (b,4) are marked with T for true as %
the last elements with that group ID (the example assumes that the next batch also starts with c, hence (c,9) becomes (c,9,F) on the next pipeline stage. Then, the marked tuples pass through the prefix scan, which appends to them (incrementally): %
a cumulative sum on their values (to facilitate sum-related operations), a cumulative sum on their valid bits (to calculate the number of tuples per group), and a cumulative sum on their recently-marked last bits (that will be used as an index to the reverse butterfly network for the round-robin effect). The first two sums are in the scope of their corresponding group IDs, and it is achieved by simply comparing their group IDs before interacting with each other according to the prefix scan topology. The next step is the additional stage in the prefix scan, where the \(n'\) entities finalise the results and also update the counts according to the previous batch. 
Here, %
(a,2) and (b,4), being selected as the tuples to carry the aggregated information for their groups a and b respectively, they replace their keys %
with the average values per group (e.g. by dividing 12 by 2 for b). The permutation indices are not altered in this case, remaining at 0 and 1 respectively, assuming this was the first batch, or that the previous batch has outputted its last aggregate result on index \(P-1\).

\section{%
EnthuseSWAG -- Sliding window aggregation}\label{swagsec}
The %
design %
of Enthuse from section \ref{gba} is adapted to perform sliding-window aggregation queries.  %
In terms of functionality, the main difference from Enthuse %
is that %
the input is processed in a windowed context. %

\begin{figure}[h!]
\centering
\includegraphics[width=0.37\textwidth, trim=0 15 0 5]{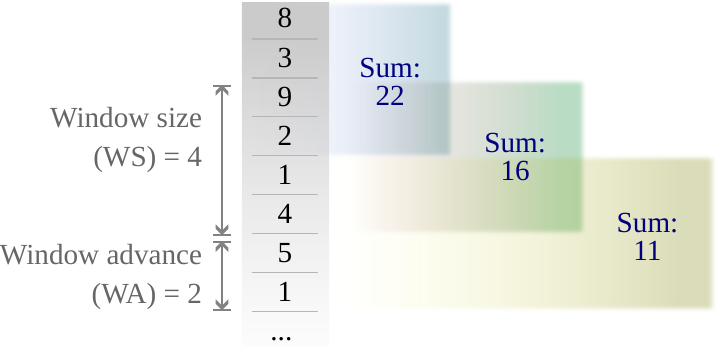} %
\caption{SWAG illustration for the \textit{sum} operator.}\label{figsw}
\vspace{-0.5 em}
\end{figure}

SWAG queries are  %
usually of the form \textit{``calculate the median of the last WS elements every WA tuples''} 
 \cite{geethakumari2023stream}, where \textit{median} can be replaced by any supported aggregate operator, \textit{WS} is the window size, and \textit{WA} is the distance between the sampled %
windows. %
Figure \ref{figsw} shows an instance %
of the \textit{average} operator, where on every 2 tuples (\textit{WA}) we find the average key %
within the last 4 tuples (\textit{WS}). In the literature, \textit{WS} is also referred to as the window range, and \textit{WA} is also referred to as the bulk size, slide or stride. In some contexts, there is a group ID associated with each key, thus, the engine can also provide per group information, as with Enthuse. %

\subsection{Supported functionality}

The proposed SWAG engine supports the operators of the base pipeline (see section \ref{subfu}), plus selection-based operators including the \textit{median}. Based on the query classification from \cite{Tangwongsan2018}, the supported types are \textit{Sum-like}, \textit{Max-like} and \textit{Sketch-like}.

In contrast to software algorithms that calculate SWAG incrementally \cite{Tangwongsan2018}, which implement the window eviction post calculation \cite{tangwongsan2017low}, here, there is no requirement for an inverse function. This is because each window is processed in its entirety, which contributes to the adaptability of the EnthuseSWAG for operators that cannot be calculated incrementally. Section \ref{disc} elaborates on the applicability of Enthuse and EnthuseSWAG to other prominent aggregation workloads.

\subsection{Design}\label{swage}

The input data are seamlessly sorted on-the-fly using a small-scale sorter (see section \ref{scard}). As opposed to the pipeline of Enthuse, the input is not expected to be sorted, since the latency of doing so is shown to be minimal %
for the window sizes of interest. Figure \ref{figswengine} introduces the pipeline of EnthuseSWAG in steps, for data arriving in batches of \(P\) tuples.

\vspace{0.2em}
\paragraph*{Step (a) Window generation} starting from the leftmost component of figure \ref{figswengine%
}, a simple buffering arrangement passes the entire window to the second component of the pipeline, which is the sorter. The buffering is used to produce WS-sized chunks (i.e. windows) while advancing by WA without the need to ``replay'' the incoming stream from an external module. This component uses a block ram-based circular buffer of size 2WS. Having the ability to store two separate windows is crucial for being able to seek back up to a window size (when window advance (WA) is at its highest possible value (WS)), as well as to efficiently support long bursts and backpressure in real systems. Due to the nature of the SWAG queries, the output rate has the ability to surpass the input rate, such as when WA$<$WS and every element needs to produce a summary (e.g. group IDs are always unique). Hence, a ``ready'' signal is %
used to enforce backpressure to the input source. %

\vspace{0.2em}
\paragraph*{Step (b) Sorting and cardinality count}a sorter is needed right before the aggregation engine to sort the window based on the \(group\) IDs, as well as the keys. Note that sorting by the keys as well is not always required, such as when the operator is commutative as with summation. %
Since 4K elements, for example, are considered moderately large as SWAG window sizes \cite{geethakumari2023stream}, the sorting can entirely be done on-the-fly instead of %
indexing larger memories. FPGA-based high-throughput sorters are thoroughly studied in literature \cite{saitoh2018high, zuluaga2016streaming, qiao2022topsort, song2016parallel}.

\vspace{0.2em}
\paragraph*{Step (c) Prefix-scans} this step remains similar as with %
the design of section \ref{gba}, but with one of the three simultaneous cumulative sums (for keys, valid bits and last bits) decoupled from the others, in order to happen first. Therefore, it now requires two prefix scan-like structures. The preceding prefix scan is on the valid bits, so that the index of each tuple within the group is known first. 

\vspace{0.2em}
\paragraph*{Step (d) Rollover and summarisation} the idea is to be able to do selection-based queries such as the median value, just by comparing this index with the group cardinality that is already-appended to all tuples by the modified sorter (section \ref{scard}). One target operator is %
\textit{min./med./max.}, which leaves the first, middle, and last tuples of each group (corresponding to the min./med./max.) as a valid aggregate result, as opposed to a single tuple for the aggregated summary of the group-by engine. %
\paragraph*{Steps (e) 2nd Prefix-scan and (f) Stream compaction} the last two steps %
reduce back to a PRRA architecture \cite{prra}, as the remaining functionality only relates to the round-robin compaction. %
The final index calculation is done by the last prefix scan in the pipeline, while the first one is now responsible for all computation-related functionality including the other two cumulative sums.

\subsection{SWAG numerical example}\label{swagex2}

\begin{figure}[h!]
\centering
\includegraphics[width=0.49\textwidth, trim=0 6 0 3]{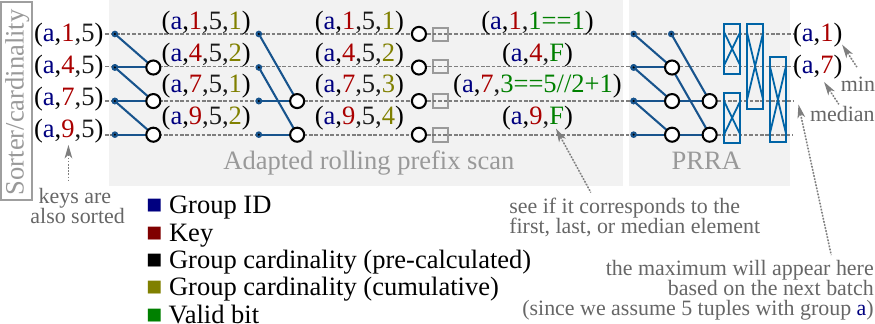} %
\caption{Colour-coded example on finding the \textit{min./med./max.} per group.}\label{fignumex2}
\vspace{-0.3 em}
\end{figure}

The numerical example of figure \ref{fignumex2} %
 demonstrates the min./med./max. selection, and %
 the expected positions of the results. %
The idea for the sorted-supplied cardinality information is to be able to perform selection %
just by comparing the cumulative group cardinality (position of a tuple inside its sorted group) with the pre-calculated one %
(section \ref{scard}).

\subsection{Sorter with group-cardinality calculation}\label{scard}

Additional logic is needed when supporting non-associative aggregation functions such as when finding the median. %
The calculation of the median value is not trivial without knowing the size of the data for which it is applied. This requires a %
modification inside the preceding sorter to be able to append the median-related information, which is the group cardinality alongside the data passed on to the aggregation engine. This functionality can conveniently be integrated into the sorter, since the sorter is provided with the entirety of the windows before flushing its sorted output to the rest of the engine. %

\begin{figure}[h!]
\centering
\includegraphics[width=0.40\textwidth, trim=0 4 0 3]{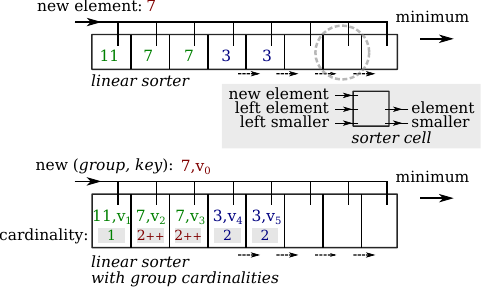} %
\caption{Adding group-cardinality support to the linear sorter.}\label{figlin}
\vspace{-0.3 em}
\end{figure}

The ability to involve median calculation within a sorter %
depends on the sorter architecture \cite{guo2001integrated}.
Our proposed solution is based on a sorter that uses the linear sorter as a building block \cite{fsorter}. The linear sorter is the hardware equivalent of insertion sort, and is able to insert a new value into an already sorted sublist \cite{fpgasort}. Each sorting cell, as depicted in figure \ref{figlin} (center), has a comparator that compares its tuple with the incoming one. According to the result of the comparison and the one of its neighbour (``left smaller''), it either adopts the new value, propagates the left value, or remains as is. 

The modification that appends %
group cardinalities exploits linear sorter's ability to have a sorted list on each cycle by comparing with every existing element. An additional comparison is done on the group subfield of the incoming tuple to determine if it is from the same group, and increments its own cardinality accordingly, as illustrated in figure \ref{figlin}. The incoming tuple then adopts the group cardinality upon insertion according to the neighbouring entity that holds the same group ID. It is guaranteed that a neighbour or its past value will have this information, since the group IDs appear clustered due to the sorting effect.

The baseline sorter of our design %
is able to provide high-throughput (multiple tuples per cycle) while supporting moderately large window sizes. The open-source %
sorter of \cite{fsorter} is specifically selected because it combines linear sorters with a merge tree (PMT \cite{song2016parallel}) for high-throughput. %
A special feature of this combination is that it is able to work efficiently in many-leaf (input list) merge mode using the same hardware, and this also applies to our new logic for resource efficiency.

\begin{figure}[h!]
\centering
\includegraphics[width=0.40\textwidth, trim=0 5 0 -2]{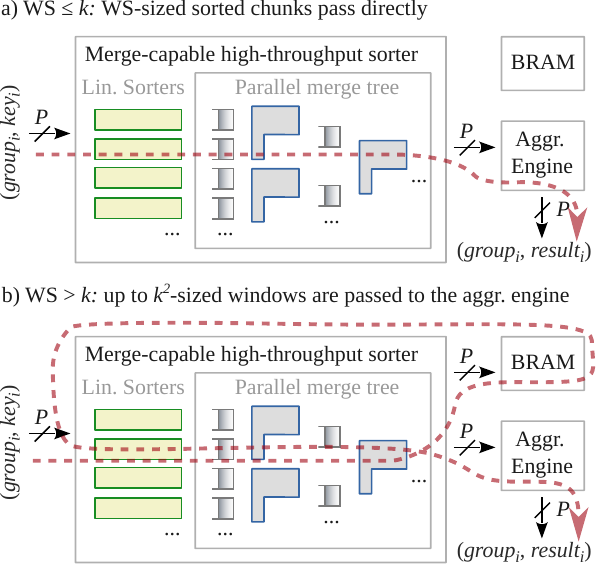} %
\caption{Data movement inside the proposed SWAG engine, according to the window size (WS) and internal size (\(k\)) of the merge-capable sorter.}\label{figdatmov}
\vspace{-0.3 em}
\end{figure}

Figure \ref{figdatmov} shows how the data moves throughout our proposed SWAG engine that uses this sorter. When the windows fit inside the sorter (\(\textit{WS}\leq k\), where \(k\) is the capacity of the sorter (i.e. the total cells of the linear sorters), the sorter works in sorting mode. In this mode, the linear sorters are flushed and then merged on-the-fly each time a window is read in its entirety. The output is then directly connected to the aggregation engine, which contains the sorted windows with each tuple having appended the new cardinality information. When \(\textit{WS}>k\), the window is sorted in chunks of size \(k\) and are stored temporarily in a BRAM buffer. The sorted chunks then reenter the sorter, but this time the sorter works in merge mode, and its linear sorters are able to perform merging by only keeping the minimum head of each sublist alongside the corresponding BRAM indices. During the merge phase, the modified sorter cells work as before, but the cardinality of incoming data is inherited from the stored values, instead of being equal to 1. Internally, this second pass makes our SWAG engine to be able to support window sizes (WS) of up to \(k^2\). %

\begin{figure}[h!]
\centering
\includegraphics[width=0.44\textwidth, trim=0 13 0 2]{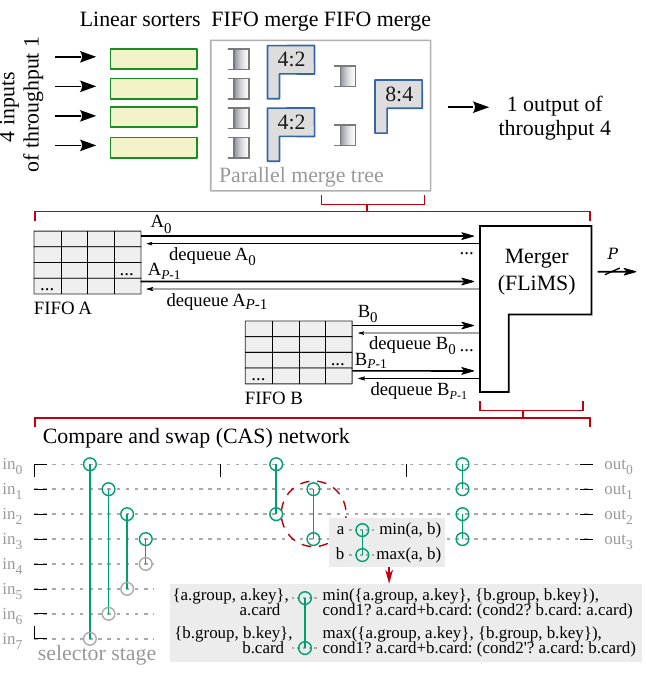} %
\caption{Adding group-cardinality support to the high-throughput sorter. %
}\label{figpmt}
\vspace{-0.3 em}
\end{figure}

An additional modification is needed to propagate the correct group cardinalities, as summarised in figure \ref{figpmt}. The parallel merge tree (PMT) needs to sum the cardinalities calculated by the linear sorters. Each FLiMS merger \cite{flimsj} of the PMT uses a network of compare-and-swap (CAS) units, which are sorters of 2-elements. The modification here is that each CAS unit uses the \textit{cond1} and \textit{cond2} conditions to determine how to update two incoming cardinalities every time they share the same group ID. The tuples carry two extra bits inside the mergers, one for denoting if the updating of their cardinalities has already happened, and one for their original source (lists A or B according to figure \ref{figpmt}). \textit{Cond1} (to sum both cardinalities) is true whenever the cardinality update has not already happened to any of the 2 inputs, and the source is different. \textit{Cond2} (to adopt the other cardinality) is true whenever the cardinality update has happened to the other tuple, and the source is either different or the same. A similar arrangement is done for the selector stage units, as they are based on the same comparison, but they also remember any merged cardinalities from the elements of the last comparison. The latter is needed to support the case when a set of same group IDs is only coming from one of the lists A or B. %

\section{Evaluation}\label{eval}

This section evaluates the implementations of the group-by aggregate and 
SWAG engines as accelerators. %
All implementations are accompanied by a sorter/ cardinality counter of capacity \(k\) of 128 elements, supporting arbitrary input sizes for the group-by engine by being applied recursively, and window sizes of up to \(\textit{WS}=16K\) for EnthuseSWAG.  %
The operator type is configured at run-time.  The implemented 
engines support multiple operators also configurable at run-time, as summarised in table \ref{tab2}.
\subsection{Experimental methodology}

The engines are evaluated as co-processors in a Linux-running environment. They are implemented as AXI peripherals, and %
the codebase is mainly written in Verilog\footnote{%
Source code: \textit{available soon at \url{https://philippos.info/nca/software}} %
}
. The entities such as the compare-and-swap units for the sorter's merger cells or the %
entities of the prefix scans are parameterisable, such as with custom data widths and functionality, and are written by hand. The structures that use those entities including the reverse butterfly network and the parallel merge tree in the sorter are generated by scripts. These are written in Python to easily infer such complex or repetitive entity topologies for arbitrary \(P\) and \(k\) (i.e. sorter size) values. %
 The resulting accelerators are evaluated on the Ultra96 development board with the Zynq UltraScale+ ZU3EG device, an MPSoC that runs Linux on its ARM A53 cores. %
This is an embedded platform, showcasing the resource-efficiency and adaptability of the proposed engines, while being scalable for higher-end platforms as well.

\begin{figure}[h!]
\centering
\includegraphics[width=0.38\textwidth, trim=0 5 0 3]{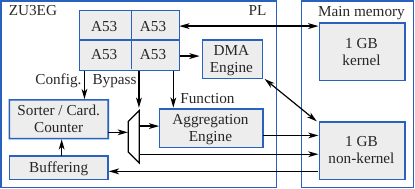}%
\caption{Experimental setup on the MPSoC platform. %
}\label{setfig}
\vspace{-0.3 em}
\end{figure}

The 2GB memory of the board is divided in two halves by the kernel;  %
the kernel-mapped part, which is optimised for cached access to the CPU, %
and the non-kernel part, which is left for the FPGA. %
The target operating frequency of the AXI-related IPs %
is 250 MHz, with an internal frequency of 125 MHz for the adapted sorter and engines, complying with the sorter's original %
design. This results in a 256-bit/cycle datapath for the engine from an 128-bit-wide HP port. %
This is relatively close to saturating the memory bandwidth of this platform \cite{manev2019unexpected}. 
Figure \ref{setfig} visualises this experimental setup. %

The data first pass through the high-throughput sorter, %
and its output is directed towards the rest of the aggregation engine. The engine is either Enthuse for group-by-aggregation %
or the corresponding part of EnthuseSWAG for sliding-windows. For the first, the sorter is the unmodified open-source IP from \cite{fsorter}, whereas the latter combines the group cardinality functionality from section \ref{scard}.   %

The core uses a direct memory access (DMA) engine to transfer the data from the main memory to the sorter-cardinality counter. The engine configuration such as \textit{function\_select} is managed by writing to a few memory-mapped registers. As soon as the data arrives at the sorter, they are immediately being sorted into chunks of up to \(k\) size. The destination of those sorted chunks are either a temporary block memory (BRAM), the main memory, or the aggregation engine. Since the proposed solutions use a merge sorter, it is a recursive operation, and the engine only reads the sorter's output on the last merging pass, hence the multiplexer. 

The sorter-cardinality counter does not currently need to output data to the main memory for the SWAG version, however, since the \(k^2\)-sized windows are deemed enough. %
Thus, the DRAM here is part of the evaluation platform and is not involved or needed in the calculation of the SWAG queries.
The handling of the data by the sorter is preceded by a simple buffering mechanism that provides the studied windows in their entirety, as described in section \ref{swage}.  %

Their performance is compared against highly-optimised software, including reference implementations from IBM for SWAG \cite{ibm}. %
The custom operators %
remain relatively simple, as they rely heavily on well-regarded %
functions and structures from C++'s standard template library (STL). 
The baseline software uses %
\textit{std::set} and \textit{std::map} to hold and update per-group information (notably 
similar performance to their hash-based \textit{unordered} counterparts, not shown), and \textit{std::nth\_element()} to calculate the medians. 
The GCC compilation flags %
are \textit{-O3 -march=native}.

\subsection{Experimental results -- Enthuse, group-by aggregation }

This part of the evaluation %
explores Enthuse %
as a standalone add-on to the existing unmodified small-scale sorter \cite{fsorter}. %
The supported operations per \(group\) ID in this implementation are: \textit{minimum, maximum, sum, count} and \textit{distinct count}. %
The engine accepts a %
data stream consisting of \((group, key)\) pairs. Each input tuple is 64-bit wide, 32-bit for each of the two fields. The pipeline throughput \(P\) is 4 tuples per cycle. 

The C++ baseline is equivalent to the functionality of the design, i.e. first sorting the data and then performing the aggregation serially. The sorting function is \textit{std::sort}, which is %
a high performing sorting baseline for general %
use, although not the absolute fastest \cite{flimsj}.
The achieved speedup for \(2^{16}\) %
tuples over the embedded processor varies between 22 to 28 according to the input distribution. This variation relates to the number of rows in the output. For instance, if all tuples have the same group ID, then the latency of interacting with the FPGA would be comparatively higher. This is because only a single tuple would be produced as an output. %

This speedup mostly comes from the sorter, as the studied CPU is already efficient for linearly skimming through a sorted stream while aggregating, also given its advantageous memory interfacing and speed. The speedup behaviour, such as for arbitrarily long streams, is inline with the sorter \cite{fsorter}, hence not elaborating with comparisons for standalone group-by. %

\subsection{Experimental results -- EnthuseSWAG, without groups}

The engine is initially evaluated for SWAG without group IDs, and a single stream of 32-bit integers is processed with high throughput. In this case, in order to saturate the 256-bit datapath, the generated aggregation pipeline and sorter-cardinality counter are for \(P=8\), and without \(group\) fields. As with the operator type, the \textit{WS} and \textit{WA} values are also configured at runtime.

EnthuseSWAG is compared against IBM's reference implementations of multiple aggregation algorithms on the platform's CPU \cite{ibm} after porting it to ARM. These include the \textit{nbfinger, amta, two\_stacks\_lite, chunked\_two\_stacks\_lite, daba\_lite} variations using the \textit{run\_bulk\_evict\_insert.py} experiment setup. The tested operations are the commonly supported \textit{sum} and \textit{min.} or \textit{max.} for data in-order, though only \textit{sum} is shown for brevity, as the performance remains relatively similar for these operators. Each kept datapoint is the minimum time from all algorithms, so that we compare with the best available software performance per window size (WS) and window advance (WA) combination.
 
\begin{figure}[h!]
\centering
\includegraphics[width=0.48\textwidth, trim=0 7 0 7]{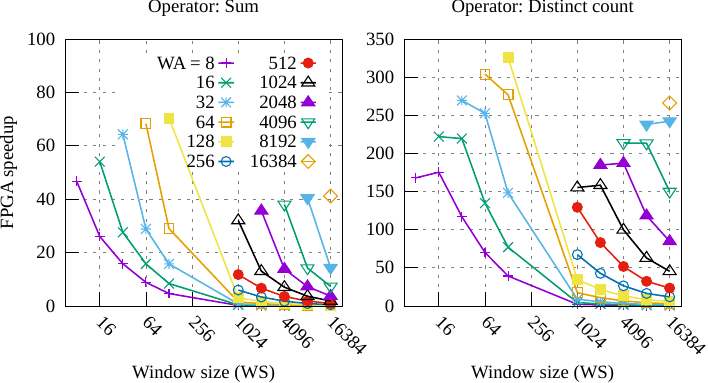}%
\caption{SWAG, without groups: engine speedup over one A53 ARM core. %
}\label{figswagng}
\vspace{-0.3 em}
\end{figure}
\begin{figure*}[h!]
\centering
\includegraphics[width=0.98\textwidth, trim=0 9 0 5]{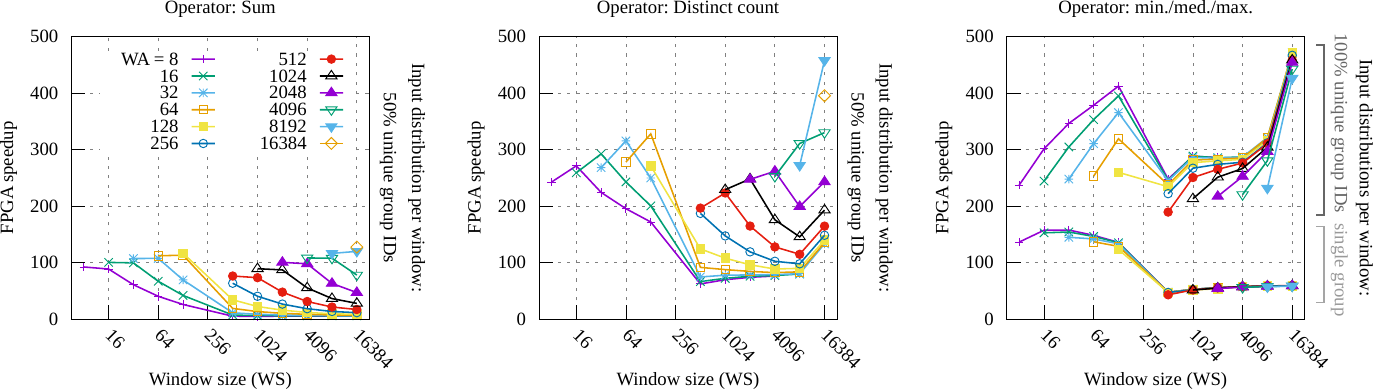} %
\caption{SWAG, with groups: engine speedup over one A53 ARM core for 3 operator types. %
}\label{figswengres}
\vspace{-0.7 em}
\end{figure*}

Figure \ref{figswagng} summarises this exploration along a custom made \textit{distinct count} routine, which is representative of more advanced queries. As it can be seen, the FPGA engine achieves up to one to two orders of magnitude speedup when WA is closer to WS. This is because the software implementations benefit from incremental computation by applying an inverse function, whereas the FPGA processes the sliding windows in their entirety. There is also a dip of performance after \(WS=k=128\) due to the requirement of an intermediate pass from the merge sorter as illustrated in figure \ref{figdatmov}b, while still providing competitive performance for the majority of WA-WS combinations, especially for  queries like \textit{distinct count}.

\subsection{Experimental results -- EnthuseSWAG, with groups}

For SWAG with groups, the evaluation also focuses on %
\textit{min./med./max}. This operator %
combines the aggregated results in the form of up to 3 tuples per group representing the {minimum}, {median} and {maximum} with the corresponding group ID in a single run. The logic behind this operator selection is to showcase a variation of algorithms that benefit from the CPU and the FPGA at different degrees. For instance, the \emph{sum} operator can be calculated incrementally, and a pointer-based solution %
can potentially 
have an advantage over the %
FPGA solution, 
as it does not have to read entire windows for smaller window updates. On the other hand, \textit{min./med./max.} is less trivial or rewarding to calculate incrementally on the CPU.

As demonstrated in figure \ref{figswengres}, the FPGA advantage of the solution %
varies for the three tested operators. %
For the majority of operator, \textit{WS} and \textit{WA} combinations, the engine provides a speedup of one to two orders of magnitude. If we ignore the dip after the internal pass is invoked (\(\textit{WS}>128\)), we can see the upward trend in the figures with respect to the window size, which can be explained by the thoughtful %
use of the on-chip memory and longer memory transfers, as opposed to the CPU's cache limitations. An exception is the combination of a high \textit{WS} and low \textit{WA} for the \textit{sum} operator where the FPGA experiences a speedup lower than 1x. This comes from the software advantage of being able to efficiently do incremental updates for such a simple operator as \textit{sum}, though according to a use case such as with 100\% unique group IDs, \textit{WS=128, WA=64} the speedup is 164x (not shown).

The studied operators also behave differently under different input distributions at least when compared to the software equivalent. The input of \textit{sum} and \textit{distinct count} operators consists of 50\% unique group IDs per group, which means every window has around \textit{WS/2} unique group IDs. The variations when changing from 100\% unique to near 0\% unique (single group ID) were relatively limited for these two operators, especially with regard to the FPGA performance. Thus, only the 50\% version is shown to conserve space. For example, for \textit{WS=16384, WA=8192} the speedup on the \textit{distinct count} operator is 317x, 420x and 476x for 100\%, 50\%, and single unique groups respectively. However, this is not the case for the operator %
\textit{min./med./max.}, hence the overlaid rightmost plots of figure \ref{figswengres} for different distributions. The output size varies according to the distribution, ranging from 1 to 3 tuples per group ID, according to the overlappings. If we select \textit{WS=16384, WA=64}, the corresponding speedups for \textit{min./med./max.} change to 59x, 239x and 472x. This is expected as the increased localities from having to operate on fewer elements can benefit software implementations when comparing the overheads of using a co-processor, while noting that the speedup is still impressive.

\subsection{Higher-end baselines}\label{higherend}
\begin{figure}[h!]
\centering
\includegraphics[width=0.43\textwidth, trim=0 7 0 5]{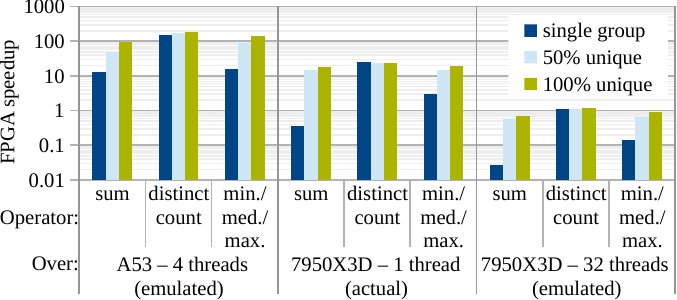} %
\caption{Comparison examples with higher-end baselines, \textit{WS=WA=16K}}\label{figopt}
\vspace{-0.3 em}
\end{figure}

Our system-level evaluation focuses on single-thread baselines %
to avoid introducing more %
variables. This is inline with reference implementations aiming at algorithmic comparison \cite{ibm}. Specifically, SWAG exhibits a performance tradeoff between incremental computation and parallelism \cite{theodorakis2020lightsaber}, %
making the selection of an oracle parallel implementation challenging. 
For the sake of completeness, a brief comparison %
is provided with an emulated parallel version on the same platform, and a %
high-end processor in figure \ref{figopt}. The parallel implementation is emulated by the single-thread version making \textit{fork()} calls. %
Then, the median wall-time is divided by the number of ``threads''. In this way, %
multi-processing inefficiencies such as memory pressure are partly reproduced, providing a lower bound for the obtained speedup. 

On the same platform, the FPGA speedup ranges from 13x to 148x over the parallel oracle version according to the input and operator. Over 7950X3D, the actual speedup for single-thread is up to 26x, while \textit{sum} with a single group is slower due to its incremental nature and data localities. %
When using 32 threads, the performance is similar at up to 1.2x, while noting the same input/operator combination outlier. Note this processor has a more advanced data source (DDR5 vs LPDDR4) and process (5 vs 16 nm). %
The FPGA power is also notably lower at 3.34W over 45W for this workload (latter influenced by DVFS%
), highlighting its appropriateness for %
edge applications, which are prominent with SWAG \cite{Tangwongsan2018}.

\subsection{Resource utilisation} \label{resul}

Table \ref{tab2} introduces a summary of the evaluated engines alongside their look-up table (LUT) and flip-flop register (FF) utilisation as reported by Vivado. %
The presented utilisation is for the corresponding encapsulating AXI peripheral. Note that slight variations can occur when further exploring optimisation directives, interfacing, register placement etc., though they remain consistent between all presented runs.

\begin{table}[h!] 
\vspace{-0.7em}%
\small
\footnotesize	
\caption{Vivado resource utilisation summary of Enthuse} 
\label{tab2}
\centering
\setstretch{0.95}
\setlength{\tabcolsep}{3pt}
\begin{tabular} {c c c | c c c }
\thead{Engine,\\input widths} & \thead{Sorter,\\input widths} &Group operators&LUT&FF\\[0.5em]%
&&&\\
N/A &\thead{unmodified \cite{fsorter},\\4\(\times\)(64-bit)}&N/A&34,101&47,122\\[0.2em]%
\thead{Group-by,\\4\(\times\)(32-bit+32-bit)}&\thead{unmodified \cite{fsorter},\\4\(\times\)(64-bit)}&\thead{min., max., sum,\\ count, dist. count}&36,653&49,544\\%
\thead{SWAG w/o groups,\\8\(\times\)(32-bit)}&\thead{Section \ref{scard}\(^1\),\\8\(\times\)(32-bit)}&\thead{min., max., sum,\\ count, dist. count,\\ min./med./max.}&29,466&41,990\\%
\thead{%
SWAG w/ groups,\\4\(\times\)(32-bit+32-bit)}&\thead{Section \ref{scard}\(^1\),\\4\(\times\)(32-bit+32-bit)}&\thead{min./med./max.}&41,963&54,437\\[-0.2em]%
\thead{SWAG w/ groups,\\4\(\times\)(32-bit+32-bit)}&\thead{Section \ref{scard}\(^1\),\\4\(\times\)(32-bit+32-bit)}&\thead{min., max., sum,\\ count, dist. count,\\ min./med./max.}&43,441&54,694\\%
\multicolumn{4}{l}{\(^1\)Also appending cardinality information.}
\end{tabular}\\
\vspace{-0.9 em}
\setstretch{1}
\end{table}

With respect to the group-by aggregate engine (Enthuse), we can see that there is about a 2.5K overhead on the number of Lookup-tables (LUTs) when becoming an add-on to an existing unmodified sorter. All engine designs here use a register per pipeline stage, which also group-by introduced a 2.4K overhead on flip-flop (FF) registers for this engine. %

The SWAG engine without group IDs consumes the least amount of resources in table \ref{tab2}. This is because of the absence of groups, which yields less resources for sorting, as it uses 32-bit values only. In terms of percentage of the ZU3 device, the LUT and FF utilisation is 42\% 30\% respectively. For all sorters/cardinality counters in the designs, the tested internal sorter capacity is steady at \(k=128\). This is despite being the only one with \(P=8\) to saturate the datapath, processing 8 \(\times\) 32-bit values per FPGA cycle.

For SWAG with groups and the \(min./med./max.\) operator, if we abstract all the SWAG functionality as an add-on to the sorter, 
the overhead totals around 8K LUTs and 7K FFs. %
These values includes the %
cardinality calculation support within the sorter for each tuple of section \ref{scard}, which makes this new hybrid sorter-cardinality counter an integral part of the design. %

In terms of block RAM, EnthuseSWAG needs to accommodate 2\(k^2\) tuples for the sliding window buffer and \(k^2\) for the merge phase of the sorter, where \(k\) is the sorter capacity. Since here \(k=128\) (in order to support a \textit{WS} of up to \(16K=k^2\)), %
this totals \(3k^2\times 64\textit{bits} = 3\textit{Mibs} \approx 85 \times 36\textit{Kibs}\), resulting in 89 BMEMs for the final SWAG engine. This is around 40\% of the %
BRAM of this device.

\subsection{Scalability}\label{scal} %
The proposed %
engines are scalable to high degrees of parallelism \(P\) and are able to saturate the bandwidth for current and future systems with a wide datapath. The total amount of data that can be produced or consumed by the engine per cycle is \(P\) times the data width. An exploration of the impact of \(P\) and the data width on the resources is conducted through out-of-context synthesis. %
The tuple (data) width is halved and used by the group ID and key fields for simplicity. The synthesis is done with Yosys \cite{wolf2016yosys} for a more generic but modern LUT6 architecture, which can serve as high-level guidelines for the resource utilisation expectations for a variety of platforms. %

\begin{figure}[h!]
\centering
\includegraphics[width=0.45\textwidth, trim=0 7 0 8]{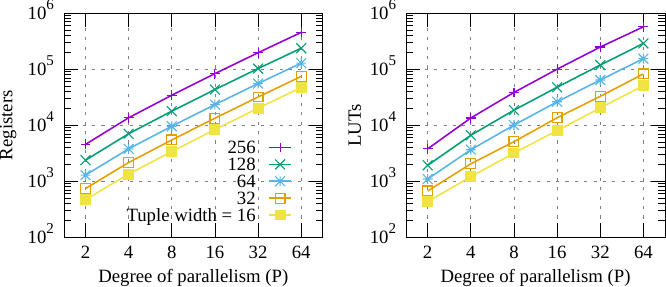}%
\caption{Synthesis results when varying tuple width and \(P\).}\label{figscala}
\end{figure}

Figure \ref{figscala} summarises the results of this exploration for EnthuseSWAG %
without the accompanying sliding window buffering and sorter. %
The sliding window buffering logic is simplistic, hence it is omitted from the scalability study, as it mainly consists of block RAM. Since %
our modifications to the sorter are relatively lightweight, the readers can also refer to the sorter's original description on its scalability \cite{fsorter}. %
For the ranges of interest, the registers and LUTs scale almost linearly with \(P\) and the tuple width. The fastest growing component is the reverse butterfly network, with a complexity of \(\Theta( P\log_2(P))\) switches (similar for the prefix scans), though the slight upward trend is not observed in these ranges yet. %
Our use case has yielded a low enough register utilisation for our purposes, but future work can also focus on retiming to considerably reduce the register utilisation when necessary.

Overall, \(P\) seems more influential to the LUT and register use than the tuple width, as it scales %
more rapidly %
for the studied ranges. Even the highest points here (\(P=64\), tuple width 256 bits) are feasible with today's datacenter FPGAs, though any complications would first relate to how easy it is to establish a sorted input stream with such a high %
throughput. 

This scalability study focuses on the aggregation part for EnthuseSWAG with groups without loss of generality, since it is the most feature-rich version of the Enthuse pipeline. The obtained curves for the other pipelines are very similar, but they are omitted for brevity.

\subsection{%
Throughput and latency}\label{throughlat}

Aggregation engines are often found as streaming accelerators, hence it can be a requirement to be able to cope with the line rate of the system, especially when the stream is continuous. As a non-blocking pipeline, %
Enthuse supports the full throughput of \(P\) elements per cycle without any backpressure for tasks such as group-by. %
Its pipeline latency of \(1+2\log_2P\) cycles is negligible for streaming data, which could be minimised further with register retiming. %

When Enthuse is extended to implement EnthuseSWAG,
it becomes a blocking design (i.e. its ``ready'' signal or equivalent is not always high). The amount of backpressure it applies on its producer mostly relates to the window advance (WA) %
in relation to its window size (WS). This is because, when \(\mathrm{\textit{WA}}<\mathrm{\textit{WS}}\) it has the potential to produce more output than input \cite{10.1145/3465998.3466014}. %
Specifically, assuming that other than the sliding window buffering there is no other %
backpressure %
from the rest of the engine %
or system, the highest throughput that can be achieved by an ideal implementation is %
\(P\mathrm{\times \textit{WS}}/\mathrm{\textit{WA}}\), where \(P\) is the degree of parallelism. %

\begin{figure}[h!]
\centering
\includegraphics[width=0.49\textwidth, trim=0 6 0 6]{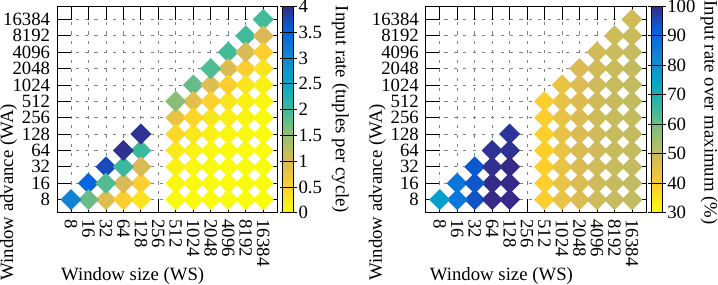}%
\caption{Final input rate influenced by backpressure, for \(P=4\).}\label{figthrough}
\vspace{-0.3 em}
\end{figure}

Figure \ref{figthrough} provides simulation results using Icarus Verilog \cite{williams2002icarus} that show the input rate by counting the number of input tuples per cycle for a relatively long input size %
 and \(P=4\). On the left plot, it can be observed that the input rate peaks near 4, especially when \(\mathrm{\textit{WS}}=\mathrm{\textit{WA}}=k=128\). This is since the sorter processes the data as they arrive, there is no buffer ``replay'' for \(\mathrm{\textit{WS}}=\mathrm{\textit{WA}}\), and any other buffering within the design such as the PMT is comparatively diminishing. After \(\mathrm{\textit{WS}}=256\), the maximum input rate is up to half of the available, since the sorter needs an internal second pass for the merging phase. On the right plot, %
 the numbers are divided by \(\mathrm{P\times \textit{WS}}/\mathrm{\textit{WA}}\) to uncover any inefficiencies, with the main observation being the halving of the throughput when \(\mathrm{\textit{WS}}>k\). %

The plots of figure \ref{figthrough} have a triangular shape, as the supported window advance cannot be greater than the window size, i.e.  \(\mathrm{\textit{WA}}\leq \mathrm{\textit{WS}}\). At \(\mathrm{\textit{WS}}=256\), there are no results due to a %
limitation within the sorter (merging at least \(P\) lists of \(k\) elements). 
If the missing \(\textit{WS}=256\) mode is needed, a build with \(k=64\) or \(256\) would suffice. Alternatively, this inherited limitation %
can be addressed at its source \cite{fsorter}, to enable splitting the sorted sublists into smaller pieces. %

\begin{figure}[h!]
\centering
\includegraphics[width=0.36\textwidth, trim=0 7 0 5]{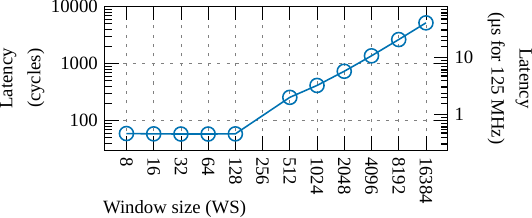}%
\caption{Simulated latency for \(P=4\), \(k=128\) and \(f_{\textit{target}} = 125\) MHz}\label{figlat}
\vspace{-0.3 em}
\end{figure}

Figure \ref{figlat} presents a similar experiment that measures the latency between the time the first tuple enters into the engine and the time of the first output. %
As shown, the \textit{WS} does not influence the latency up to \(\textit{WS}=128\), since the linear sorters in the sorter have a fixed latency (\(k/P\)) for an input to be extracted. For \(\textit{WS}>k\) the latency scales linearly with \textit{WS}, as the sorted windows are merged with a fixed rate \(P\), and the merge phase is compulsory. %
\textit{WA} does not impact the latency.

These results are generalisable to all queries using the SWAG architecture and tuple widths, because the output format does not influence the backpressure as long as up to \(P\) tuples are accepted at the output of the engine. %

\section{Related work}\label{rlw}\label{finalcomp}

There is a moderate amount of research on database acceleration focusing on analytics \cite{fang2020memory}. %
Similar to earlier works that accelerated joins \cite{casper2014hardware,owaida2017centaur}, there are tradeoffs when selecting between a sort and an indexing-based solution \cite{moghaddamfar2023study,absalyamov2016fpga}, each with different adaptation challenges
 \cite{matas2022automated, eryilmaz2021fpga}. For instance, even with a sorter-based pipeline, supporting arbitrary data in a join accelerator caused stalls and increased its memory requirements \cite{casper2014hardware}. This relates to the existence of duplicates of unknown cardinality, which is somewhat reminiscent of median support, which affects the complexity of SWAG designs. %

\begin{table*}[h!] 
\vspace{-0.7em}%
\footnotesize	
\caption{SWAG engine implementation comparison %
} 
\label{tab3}
\centering
\setstretch{1.2}
\setlength{\tabcolsep}{1.5pt}%
\vspace{-0.5 em}
\begin{tabular} {r| c ;{2pt/1pt}c ;{2pt/1pt}c;
{2pt/1pt}c ;{2pt/1pt}c%
;{2pt/1pt}c;{2pt/1pt}c;{2pt/1pt}c
}
\multicolumn{1}{c|}{}
&\multicolumn{1}{c|}{DFE \cite{8977927}}
&\multicolumn{3}{c|}{StreamZip \cite{%
geethakumari2023stream}}&\multicolumn{2}{c|}{\emph{EnthuseSWAG implementations}}&\multicolumn{1}{c|}{Mueller et al. \cite{mueller2009data}}&Oge et al. \cite{6726885}\\ \cline{2-9}
{SWAG context}&\multicolumn{4}{c;{2pt/1pt}}{{w/ groups (window per group)}}& {\scriptsize w/ groups (single window)}&\multicolumn{2}{c;{2pt/1pt}}{ w/o groups}& time-based\\[0em]\cdashline{2-9}[2pt/1pt]
{Group operators}&\multicolumn{2}{c;{2pt/1pt}}{\scriptsize {min., med., avg, max.}}&\multicolumn{2}{c;{2pt/1pt}}{min.}&\multicolumn{2}{c;{2pt/1pt}}{{min., max., sum, count, dist. count, min./med./max.}}&med.&\scriptsize min., max., sum, count\\[0em]\cdashline{2-9}[2pt/1pt]
{Max. input rate}&\multicolumn{4}{c;{2pt/1pt}}{0.5}&4&8&\multicolumn{2}{c}{1}\\[0em]\cdashline{2-9}[2pt/1pt]
Line rate&\multicolumn{4}{c;{2pt/1pt}}{{128-bit@78.13 MHz}}&\multicolumn{2}{c;{2pt/1pt}}{128-bit@250 MHz}&32-bit@100 MHz&128-bit@200 MHz\\[0em]\cdashline{2-9}[2pt/1pt]
{FPGA device}&\multicolumn{4}{c;{2pt/1pt}}{{Intel 5SGXAB}}&\multicolumn{2}{c;{2pt/1pt}}{AMD ZU3EG}&XC2VP30&XC6VLX240T\\[0em]\cdashline{2-9}[2pt/1pt]
{Logic (K)}& 86\(^1\)%
&104\(^1\) &219\(^1\)&308\(^1\)&{43 LUT, 55 FF}&{29 LUT, 42 FF}&6.4 LUT, 3.8 FF& \(<\)1.5 LUT,  \(<\)3 FF\\[0em]\cdashline{2-9}[2pt/1pt]
{BRAM (M20K)}&{1136}%
&{1584}&1796&{2139}&89 RAMB36E2 & 92.5 RAMB36E2 & 0 &17 RAMB36 \\[0em]\cdashline{2-9}[2pt/1pt]
{BRAM (Mbits)}&22.4%
&31.2&35.4&42.1&{3.3}&3.5&0&0.6\\ \cdashline{2-9}[2pt/1pt]
{DSPs}&\multicolumn{3}{c;{2pt/1pt}}{0}&165&\multicolumn{4}{c}{0}\\[-0em]\cdashline{2-9}[2pt/1pt]
{Tuple width}&\multicolumn{4}{c;{2pt/1pt}}{\thead{(32-bit time.,)\vspace{-0.4em}\\ 32-bit group ID,\vspace{-0.4em}\\ 64-bit  key}}&\thead{32-bit group ID\vspace{-0.4em},\\ 32-bit key}&\multicolumn{2}{c;{2pt/1pt}}{32-bit key}&\thead{(32-bit time.,)\vspace{-0.4em}\\  32-bit group ID,\vspace{-0.4em}\\ 2\(\times\)32-bit key}\\[-0.5em] \cdashline{2-9}[2pt/1pt]
{Key format}&\multicolumn{2}{c;{2pt/1pt}}{{\scriptsize fixed-point}}&{\scriptsize fl.-point} &{\scriptsize fl.-point (lossy)}&\multicolumn{2}{c;{2pt/1pt}}{integer, fl.-point\(^2\)}&\multicolumn{2}{c}{integer}\\[0em]\cdashline{2-9}[2pt/1pt]
WS range\(^3\)&\multicolumn{4}{c;{2pt/1pt}}{{\{\(64,128, ..., 4K\)\}}}&{\scriptsize\(\{4,8, ..., 16K\}\)  \(-\{256\}\)}&{\scriptsize	\(\{8, 16, ..., 16K\}-\{256, 512\}\) }&\{8\}&\scriptsize WA\(\times\)\{64, 128, ..., 4\(K\)\}\\[0em]\cdashline{2-9}[2pt/1pt]
WA range&\multicolumn{4}{c;{2pt/1pt}}{{\{\(1,2,4, ..., 4K\)\}}}&{\scriptsize \(\{4,8, ...,16K\)\}}&{\scriptsize \(\{8, 16, ...,16K\)\}}&\{1\}&N/A\\[0em]\cdashline{2-9}[2pt/1pt]
{Off-chip memory}&DRAM&\multicolumn{3}{c;{2pt/1pt}}{{DRAM \& 72 MB QDR-SRAM}}&\multicolumn{4}{c}{0}\\

\multicolumn{4}{l}{}\\[-1em]
\multicolumn{9}{l}{\(^1\)Intel's ALM is not directly comparable, but implies more logic per unit. %
\(^2\)All operators except summation also support IEEE-754 floating points.}\\[-0.4em]
\multicolumn{9}{l}{\(^3\)For \textit{WS} and \textit{WA} here, \(K\) denotes kibi (1024), not kilo (1000).}\\
\end{tabular}
\vspace{-1 em}
\setstretch{1}
\end{table*}

\subsection{Group-by aggregation}

Group-by aggregate operators %
have usually been a part of larger analytics designs \cite{lisa2018column}. This is due to the bandwidth limitations \cite{eryilmaz2021fpga}, as with co-processors' reduced performance when it comes to accessing main memory. This makes stand-alone solutions more challenging to become %
practical. 

A group-by aggregate-focused work demonstrated a speedup of 10x over the CPU \cite{absalyamov2016fpga}, but used multi-tasking, hashes and off-chip memory on a discontinued platform. 
A recent group-by-aggregator uses sorting to accelerate these operations \cite{xue2024fpga}, though it implements an SQL-based data management system as a whole, which is still inline with the observation on the scarcity of out-of-context group-by-aggregate accelerators.

\begin{figure}[h!]
\centering
\includegraphics[width=0.43\textwidth, trim=0 10 0 7]{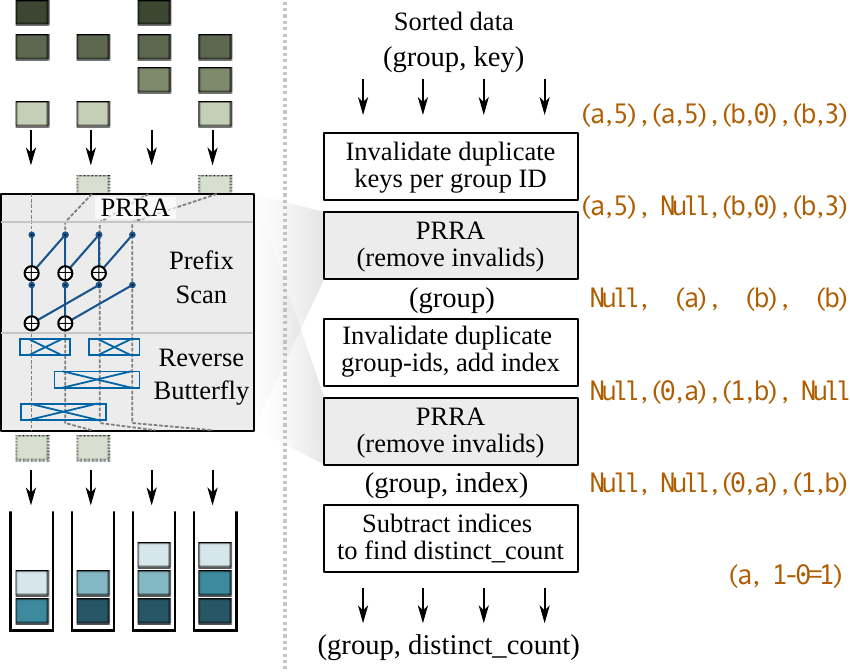} %
\caption{Modular design of the distinct count query (right) \cite{fsorter} with example data in orange, based on the PRRA implementation (left) \cite{prra}.}\label{figprra0}
\end{figure}

Despite this, the acceleration of specific group-by queries is useful in certain industrial contexts, such as \textit {distinct count per group} for customer analytics \cite{fsorter}. This operator is supported by the presented engine, but can also be implemented by combining two PRRAs for intermediate sparsity removal, as shown in figure \ref{figprra0}.

When we compare the engine's utilisation (2.55K LUT, 2.42K FF, see table \ref{tab2}) to the unmodified PRRA (1.26K LUT, 2.29K FF \cite{prra}), it is easy to conclude that %
it uses at least 1.9x fewer FFs than any solution with two PRRAs%
, and similar LUTs. %
This can be explained with the help of a simple %
model: when we abstract the node complexity to 1 for the butterfly network switches and all other entities, the complexity of the example PRRA-based engine of figure \ref{figprra0} decreases from \(3P + 2 \times \mathrm{PRRA}\) to \(2P + 1\times \mathrm{PRRA} = 2P\times \log_2(P)+P+1\) entities. This is while supporting multiple operators reconfigurable at runtime. %
In this instance, our proposed group-by engine reduces the best known resource complexity for performing \textit{distinct count per group}, while being highly adaptable for other such advanced operators.

\subsection{SWAG engines}\label{stoa}

On the other hand, SWAG engines exist stand-alone, making comparisons more direct. Table \ref{tab3} presents a comprehensive quantitative comparison on how the proposed approach compares to the state-of-the-art and representative works. 

Starting from the left of table \ref{tab3}, the comparison begins with SWAG with groups. State-of-the-art FPGA solutions provide a notable speedup over prior work \cite{8977927}, but still operate on up to a single tuple %
per cycle \cite{%
geethakumari2023stream, gong2021shuntflowplus}. They often involve the implementation of %
a multi-level memory hierarchy based on off-chip memories%
, which is reflected in the complexity of the design. In contrast, our proposed engine achieves to be evaluated on a resource-constrained embedded platform (no loss of generality to high-end FPGAs). %

An important %
difference between the presented SWAG engine and the selection of competing engines for SWAG with groups is that there is a single sliding window, as opposed to a window per group (elaborated in section \ref{disc}).
In this case, our design
uses {no DRAM} and a  {fraction of the BRAM} at 3.3 versus 22.4 Mbits for the most lightweight of the alternatives %
\cite{8977927}. %
The competing engines have a target frequency of 150 MHz, but at 125 MHz internally, it is able to saturate a {line rate over 3 times higher}. %
In terms of throughput, the competing variations in the table are essentially exchanging %
resources and functionality for %
performance (not shown). For example, DFE utilises around \(1/7\) of the input rate, even when \textit{WS$=$WA}. %
The compression in StreamZip \cite{geethakumari2023stream} causes the throughput to scale differently according to the relationship between \textit{WS} and \textit{WA}, which can be beneficial in certain applications with a low \textit{WA}. %

In terms of latency, the proposal is always {faster for $WS\leq128$} at less than 1 $\mu$s. For %
\textit{WS$=$4K}, %
the StreamZip variations have a latency advantage of near 1 $\mu$s versus around 10 $\mu$s for the proposed solution. %
Though, it becomes substantially higher for smaller \textit{WA}s, whereas the latency of our engine is {not influenced by \textit{WA}}. %
The maximum throughput of StreamZip and similar works \cite{8977927} is 70 million tuples per second (MT/s), whilst our approach is at {500 MT/s}. This is %
while supporting {4x higher \textit{WS}s and \textit{WA}s}, %
and supporting {multiple operators}. %

When it comes to %
SWAG without groups, Mueller et al. propose a sliding median operator, but was more of a proof of a generalisable concept, also given the state of the technology of the day %
\cite{mueller2009data}. The sliding window size is fixed to only 8, hence the reduced need for BRAM and DRAM. The median calculation is based on shift registers and a sorting network, which, even with newer technology, do not have the scalability of more sophisticated designs. For instance, the merge-capable linear sorters \cite{fsorter} enable a practical use with WS of thousands of elements in the presented engines. It operates at a single input per cycle, though parallelism is demonstrated for multiple workers operating on independent streams. A single worker has a throughput of 100 MT/s, while our approach reaches 1 GT/s.

Finally, a time-based SWAG engine is indicatively compared with the proposed %
engines due to its similarity in certain aspects \cite{6726885}. While groups are supported in its input similar to count-based SWAG with groups, the results are only for a single group. %
It is the most space efficient design, but lacks the high throughput and more advanced operators. It focuses on achieving a steady latency and throughput  for different WS and WA ratios, but each of those combinations requires a separate bitstream, since it is not runtime configurable. %
 At a single tuple per cycle, its throughput is 200 MT/s.

Overall, EnthuseSWAG's %
contributions stand between SWAG \emph{with groups} and \emph{without groups} including, enabling big data processing with more advanced operators for simple devices. It is worth noting that the comparison is, hence, relatively diverse, as it features partly different systems and operator combinations, and further customisation is possible. %

\section{Discussion and Future work}\label{disc}

The engines are highly adaptable and parameterisable.
\vspace{0.2em}
\paragraph*{Other operators} The selection of operators in the aggregation pipeline can easily incorporate other associative operators such as \textit{multiplication}, \textit{bitwise exclusive-OR} etc. This can be achieved by including the corresponding logic inside the \(n\) and \(n'\) entities. 
The average operation is %
not explored beyond simulation %
to avoid shifting the focus to %
floating point division optimisation. This is without loss of generality, only requiring \(P\) dividers, 
usually consuming a few hundred LUT/FFs according to the %
use case, and without a perceptible timing overhead \cite{divfor}.

Similarly, other types of averaging are possible by combining simultaneous associative operators together at the \(n'\) stage. As an example, combining \textit{multiplication} and \textit{count} with a root circuit, can provide the \textit{geometric mean}.
Another instance is with the selection mechanism inside EnthuseSWAP, which can %
be generalised to support the percentile operator %
to select %
arbitrary %
percentiles based on 
the pre-calculated cardinality. Other selection queries could use more specialised criteria. Future applications could exploit the additional information in the engine, such as the order of the sorted data, which has already been used in the \textit{distinct count} algorithm (step \((c)\) in section \ref{gba}) to essentially make it an associative operation. 

\vspace{0.2em}
\paragraph*{SWAG context} In the comparison of section \ref{stoa}, a unique aspect for EnthuseSWAG with groups is the single window, while still providing per group statistics. Per-group window SWAG, although supported, might be less trivial to efficiently exploit the presented architecture, as the sorting advantage might diminish when ``recycling'' more data on each window advance. Thus, it could be seen as an approximation for when the task involves a higher number of groups and there is a need for multiple windows. In practice, keeping live statistics for all groups may not be necessary \cite{6726885}, and a simple filtering step to isolate the group IDs would still dedicate the whole window to the desired group. %

Yet, the notion of having a common window for grouped data is widely useful for other SWAG %
tasks. One such case is for supporting more complex SWAG queries without groups, where the group field can carry the value instead, for operating in the context of the values. This includes the \textit{mincount} or \textit{maxcount} operator (i.e. counting the instances of the minimum or maximum value within the window) \cite{tangwongsan2021order}, which can be implemented as a group-by query where the \textit{count} operator is only activated on the first or last group, corresponding the minimum or maximum respectively.

It is also especially useful when using a time-based window eviction policy, such as when summarising statistics for the last few seconds \cite{6726885}. The presented aggregation pipeline, as well as the sorter extensions for cardinality calculation actually support both count-based and time-based SWAG, since each cell automatically carries an oscillating counter %
to distinguish between arbitrarily-sized windows, assuming \textit{WS} is sufficient. This essentially means that the sliding-window mechanism can be adjusted accordingly for the whole engine to support different %
window contexts. Time-based aggregation requires minimal modification at step \((a)\) in section \ref{swage} to produce windows based on a timestamp. The present evaluation can be considered representative for time-based SWAG, if the order of the data has a time notion, such as when each new tuple is generated on a fixed interval.

\vspace{0.2em}
\paragraph*{Specialisation and optimisation} The presented engines %
can be simplified further according to the desired operator selection. If only associative operators (e.g. \textit{sum}) are needed, then the requirement for the sorted input is only with respect to the group. For SWAG, this essentially means that the comparison width of the small-scale sorter can be constrained to that of the \textit{group} to minimise its complexity. For non-SWAG, this would also be reflected in the accompanying sorter, or in software equivalents that provide the sorted stream.
If selection operators such as the median are not present, then the cardinality logic (section \ref{scard}) is not needed. Hence, the disjoint prefix-scan steps can also be merged back. In this scenario, alternative sorters would also suffice for SWAG. When this holds and the input does not carry group information, then a sorter is not needed at all. %

As demonstrated in the study of section \ref{resul} are scalable for wide datapaths.
The throughput and latency behaviour of EnthuseSWAG %
can be optimised further. %
When a higher throughput is needed for large windows, %
a wider \(P\) can be used, or  two sorters can be combined and de-multiplexed %
using %
 rate converters. %
A lower latency can be achieved by shortening the linear sorter modules, such as by using lower \(k\) values for \(\textit{WS}\leq k\), or a higher \(P\) for \(\textit{WS}> k\). 

\vspace{0.2em}

Other future work includes the automation of synthesising %
customised aggregation engines, such with a %
high-level synthesis (HLS) library. 
It could also be explored as a functional unit inside CPUs to work as %
 dedicated SIMD instructions \cite{cui2023risc}. %

\section{Conclusions}\label{con}

The increasing memory datapath widths are becoming increasingly challenging to exploit with existing programming techniques and models. This includes aggregation, a naturally parallel problem. %
Enthuse, a modular and highly-adaptable aggregation pipeline is presented that uses prefix-scan and reverse butterfly topologies to perform a wide range of aggregation tasks with high-throughput. This design improves the theoretical and synthesis-based circuit complexity for specialised group-by applications such as for counting unique elements per group. 
It is then adapted to EnthuseSWAG, to work as a competitive sliding window aggregation engine by
incorporating a modified %
sorter that sorts each incoming buffered window on chip. This includes %
lightweight algorithmic modifications %
that append group cardinality information to the incoming tuples to assist with median calculation. %
The resulting designs %
have low hardware overheads, while often supporting a superset of the functionality of alternatives. EnthuseSWAG exhibits up to hundreds of times of speedup when compared to CPU-only approaches for the same task on our embedded platform. It is also able to process multiples of the input rate of the state-of-the-art. For SWAG with groups, the competitiveness with prior FPGA work %
is achieved under the practical assumption of a single window. %
This is already implied for tasks such as SWAG without groups, where EnthuseSWAG processes multiple tuples per cycle without multiple workers. %
The highly-optimised on-chip pipelines eliminate the need for off-chip memories, %
and reduce the latency for relatively advanced aggregation tasks. %

\section*{Acknowledgements}
\footnotesize
The %
support of  UK EPSRC (grant number UKRI256, EP/V028251/1, EP/N031768/1, EP/S030069/1, and EP/X036006/1) is gratefully acknowledged. %
Prof David Gregg has provided valuable %
feedback on an early draft. %

\AtNextBibliography{\fontsize{8.2}{8.2}\selectfont} %
\printbibliography

\vspace{-4em}
\begin{IEEEbiography}[{\includegraphics[width=1in,height=1.25in,clip,keepaspectratio]{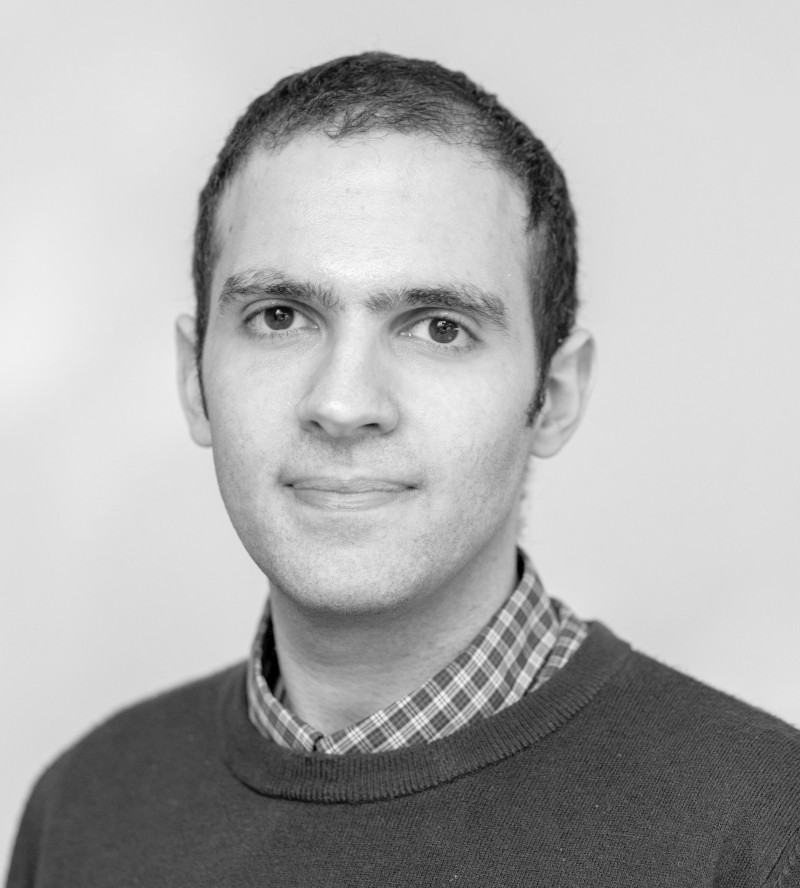}}]{Philippos Papaphilippou}
is a Lecturer in Electrical and Electronic Engineering at the University of Southampton. His doctoral studies at Imperial College London (2021) were funded by dunnhumby/Tesco to research novel accelerators for customer analytics. Other industrial engagements included a 3-month contractorship with Microsoft, and a 2-year senior CPU architect position at Huawei. His research topics include parallel algorithms, multi-processors, FPGAs and data science.

\end{IEEEbiography}
\vspace{-4em}

\begin{IEEEbiography}[{\includegraphics[width=1in,height=1.25in,clip,keepaspectratio]{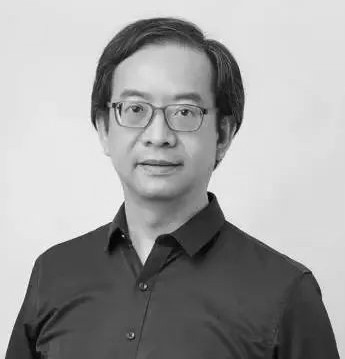}}]{Wayne Luk}
is a Professor of Computer Engineering at Imperial College London. He leads the Programming Languages and Systems Section, and the Custom Computing Research Group at the Department of Computing. He is a Fellow
of the Royal Academy of Engineering, the IEEE, and the BCS.
\end{IEEEbiography}

\end{document}